\documentclass[lettersize,journal]{IEEEtran}
\usepackage[caption=false,font=normalsize,labelfont=sf,textfont=sf]{subfig}
\usepackage{graphicx} 
\usepackage{amsmath}
\usepackage{amssymb}
\usepackage{multirow}
\usepackage{booktabs}
\usepackage{cite}

\usepackage{xcolor}
\usepackage{ulem}

\usepackage{pgfplots}
\pgfplotsset{compat=1.18} 
\usepackage{amsmath}

\usepackage{hyperref}
\hypersetup{
    colorlinks=true,
    linkcolor=blue,
    filecolor=magenta,      
    urlcolor=blue,
}

\usepackage{enumitem}
\usepackage{xurl}

\title{Practical Deep Learning-Based Block Demapping for mmWave MIMO Systems}

\author{Syed Asrar ul haq, Jesus O. Lacruz, Joerg Widmer, and Sumit J. Darak

\thanks{ Syed Asrar ul haq, and Sumit J. Darak are with Electronics and Communications Department, IIIT-Delhi, India-110020 (e-mail: \{syedh,sumit\}@iiitd.ac.in)

Jesus O. Lacruz, and Joerg Widmer are with IMDEA Networks Institute, Madrid, Spain (email: \{jesusomar.lacruz, joerg.widmer\}@imdea.org)
}
}

\begin{document}

\maketitle

\begin{abstract}
    Deep learning is expected to play a central role in 6G systems, by enabling data-driven physical layer designs that can adapt to practical wireless impairments. This is particularly important for millimetre-wave (mmWave) multiple-input multiple-output (MIMO) systems, which are strongly affected by residual channel distortions, interference, and hardware non-idealities that are difficult to capture with analytical models. In this paper, we propose a deep learning-based soft demapper that performs block-wise processing of received symbols instead of conventional symbol-wise detection. The demapper employs a one-dimensional convolutional neural network to learn temporal structure induced by practical propagation and hardware effects, while producing soft bit outputs compatible with soft-decision decoding stages. The proposed approach is trained and evaluated for a 4$\times$4 wideband mmWave MIMO system at 60~GHz, using an over-the-air dataset collected across multiple days and settings. Experimental results demonstrate consistent improvements over a conventional soft-decision demapper, including up to a 2.6$\times$ reduction in bit error rate for 16QAM and up to a 17.6$\times$ reduction for 64QAM under real over-the-air mmWave MIMO conditions. The proposed demapper is further implemented on a Zynq RFSoC platform using a reconfigurable convolution engine and fixed-point arithmetic, achieving an approximately 34.3$\times$ speedup compared to a processor-based implementation, and demonstrating that learning-based demapping can be deployed efficiently on resource-constrained edge hardware.
\end{abstract}

\section{Introduction}

Deep learning is increasingly being explored as a tool to enhance the physical layer of wireless communication systems, motivated by the growing complexity of modern air interfaces and the limitations of purely model-based designs~\cite{sur3,6gSur,sur1}. In next-generation networks, wide bandwidths, high carrier frequencies, and dense antenna configurations expose receiver 
to impairments that are difficult to capture accurately using analytical models alone. Data-driven approaches have therefore attracted attention as a way of complementing conventional signal processing by learning directly from observed waveforms. As a result, a growing body of research in both academia and industry has examined how neural networks can be integrated into the physical layer to improve robustness and adaptability under practical operating conditions~\cite{saadArtificialGeneralIntelligence2025,ai6g,aiSurvey2,deepsig_mission_critical,hoydis6GAINativeAir2021,omnisig}.

Learning-based physical layer (PHY) design has been explored at different levels
. One line of work considers end-to-end transceiver learning, where the entire transmitter–receiver chain is trained as an autoencoder that adapts to channel and hardware characteristics \cite{shea,neuralReceiver}.
Although such methods demonstrate the potential of deep learning to jointly optimise all processing stages, their deployment in practical systems is challenging, as they require substantial changes to standardised PHY and offer limited support for backward compatibility or fallback to conventional processing. To address these limitations, learning has also been introduced across multiple receiver blocks, where neural networks jointly optimise functions such as channel estimation, equalisation, and demapping within a single learned module, while still preserving a high-level receiver structure \cite{imdea_1,deeprx}.
A more modular line of work focuses on learning-based enhancement of individual receiver components, in which a single processing block is replaced or augmented by a learned model, while other stages remain model-based. Representative examples include learning-based channel estimation \cite{C_2,tcasai,9789137}, equalisation \cite{equalization}, and demapping \cite{demapper1,demapper2,demapper3,llring}.
This 
approach allows targeted performance improvements with minimal disruption to the existing signal processing chain, simplifies integration into legacy systems, and ensures backward compatibility with standardised 
designs. 

Much of the 
progress in learning-based PHY design relies on synthetic datasets generated through link-level simulators. While simulation-based evaluation is essential for controlled experimentation and rapid development, simulated channels cannot fully capture the behaviour of real radio links. Practical wireless systems are affected by hardware drift, nonlinearities, 
interference, and site-specific scattering, all of which are difficult to model 
in simulation \cite{otaNeed1,otaNeed2}. As a result, models trained exclusively on synthetic data may exhibit strong performance in controlled settings, yet fail to generalise when deployed in real environments. This gap underscores the importance of 
over-the-air (OTA) measurements. However, collecting large OTA datasets with raw IQ samples and reliable ground-truth labels is time-consuming, and only a limited number of such datasets are publicly available, which restricts reproducibility
.

Few works have incorporated hardware measurements into the evaluation of learning-based receivers ~\cite{9789137, hil,riku2025,omnisig}, but important limitations remain. \cite{9789137} demonstrated OTA training for learning-based channel estimation, but the achieved performance remained below that of the best conventional model because the conventional receiver outputs were used as training targets. Hardware-in-the-loop approaches capture some RF impairments using channel emulators, while avoiding transmission through a real propagation environment \cite{hil}.
Similarly, many works report OTA validation only at test time, while training 
predominantly on synthetic limitting generalisation when real deployment conditions differ from the simulated scenarios~\cite{riku2025}. End-to-end learned transceivers have also been demonstrated using real hardware, but their high computational complexity and reduced modularity make integration into practical systems challenging \cite{omnisig}. Together, these observations 
highlight the need for 
learning-based PHY designs supported by large, high-quality OTA datasets, particularly at frequencies and bandwidths where conventional channel models struggle to reflect real-world behaviour.

These challenges are amplified in millimetre-wave MIMO systems, making over-the-air evaluation particularly critical. Operation at high carrier frequencies and wide bandwidths increases sensitivity to phase noise, front-end nonlinearities, and timing inaccuracies.
In wideband systems, frequency-selective fading and residual impairments can exhibit noticeable intra-block variability, with post-equalisation distortions influenced by both temporal and spatial characteristics of the channel~\cite{mmWaveChallenges1}. These effects interact in ways that are difficult to capture accurately with analytical models having a pronounced impact on receiver performance. As a result, mmWave MIMO constitutes a stringent test case for learning-based physical layer techniques, where validation using measured over-the-air signals is essential to assess robustness and practical viability.

A promising point of integration for learning within the receiver is the demapping stage, where soft information is generated for each received symbol. The demapper 
is sensitive to residual distortions that persist after equalisation. Existing learning-based demapper designs, however, exhibit important limitations when considered for practical mmWave MIMO receivers. Approaches based on convolutional self-attention for multi-user MIMO introduce substantial computational complexity, which limits their suitability for low-latency and resource-constrained receivers~\cite{demapper3}. Other designs rely on multilayer perceptrons that operate on individual symbols, restricting the model to symbol-wise processing and preventing it from exploiting temporal or spatial structure in the 
waveform~\cite{demapper2,llring}. Classical demappers share this limitation, as they rely on analytical decision rules and parametric noise models
, offering no mechanism to capture short-range dependencies across time or across spatial paths that commonly arise in practical mmWave MIMO systems. Importantly, these demapping approaches have not been validated using large-scale OTA measurements in wideband mmWave MIMO settings, where residual hardware impairments and propagation effects introduce structured, non-Gaussian distortions
, leaving open questions regarding their robustness in practice.

To address these limitations, namely reliance on symbol-wise processing, high computational complexity of existing learning-based solutions, and lack of validation using OTA measurements, this paper proposes a deep learning-based soft demapper that performs block-wise processing in MIMO systems. The demapper processes a block of demodulated received symbols and generates soft bit outputs compatible with standard decoding stages, allowing it to exploit temporal structure that symbol-wise methods cannot capture. The demapper operates independently on each spatial stream and is not tied to a fixed antenna configuration, enabling scalability to different MIMO orders without architectural changes. The proposed demapper is also implemented on an FPGA-based system-on-chip platform to experimentally verify its low computational complexity and suitability for edge deployment.

The contributions of this work are as follows
\begin{enumerate}[leftmargin=*]
 
    \item We propose a deep learning–based soft demapper that operates on blocks of received symbols, improving bit-level reliability while preserving the modular structure of the receiver chain. The proposed demapper is flexible with respect to the MIMO configuration and can be applied to different antenna orders without architectural modification, making it a scalable solution for practical systems.

    \item We demonstrate consistent gains over a conventional soft-decision demapper through extensive evaluation on real over-the-air measurements, including up to a 2.6$\times$ reduction in bit error rate for 16QAM and up to a 17.6$\times$ reduction for 64QAM under practical operating conditions. 

    \item We verify that the trained demapper generalises well to different measurement days and deployment environments without retraining, indicating robustness to changes in channel conditions.

    \item We conduct a systematic ablation study covering model size, side information, and architectural choices, identifying configurations that offer favourable trade-offs between performance and implementation complexity.

    \item We collect and release a large-scale OTA dataset for wideband single-carrier mmWave MIMO systems, comprising raw IQ samples, channel estimates, and aligned bit labels measured with a MIMO testbed at 60 GHz. To the best of our knowledge, this is the first open dataset of its kind.

    \item {\color{black}We implement the proposed DL based demapper on an FPGA-based platform and demonstrate its suitability for edge deployment, achieving 18.4$\times$ acceleration compared to a processor based implementation, and up to 34.3$\times$ acceleration using pipelined parallel processing with fixed point implementation while using 40\% fewer multipliers.}

\end{enumerate}


\section{System Design}\label{sec:sysDesign}

\begin{figure}
    \centering
    \includegraphics[width=\linewidth]{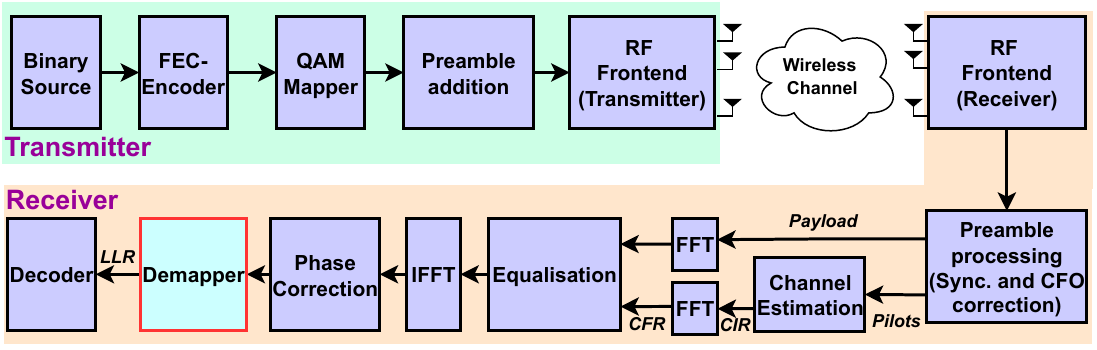}
    \caption{\small Block diagram of a wideband MIMO transceiver. The proposed DL-based demapper replaces the conventional demapper, while all other receiver blocks remain unchanged.} 
    \label{fig:blockDiag}
\end{figure}

We consider a wideband MIMO communication system (Fig.~\ref{fig:blockDiag}). The formulation 
applies to both single-carrier and multi-carrier (e.g., OFDM) schemes.
At the transmitter, a binary sequence is encoded using a forward error correction code. The encoded bits are mapped to complex constellation symbols using an $M$-QAM modulation scheme. Known training sequences are inserted within the frame to enable packet detection, synchronisation (time, frequency), and channel estimation at the receiver. The transmitted frame therefore consists of a preamble carrying training symbols, followed by payload data symbols. Depending on the waveform, these symbols may correspond to time-domain samples (single-carrier) or frequency-domain subcarriers (OFDM). 

The transmitted frame propagates through a 
wireless MIMO channel affected by additive noise as well as hardware impairments such as frequency offset, phase noise, and nonlinear distortions introduced by the transmitter and receiver front-end. The received baseband signal can be expressed as
\begin{equation}
\mathbf{y}[n] = \sum_{l=0}^{L-1} \mathbf{H}[l]\mathbf{x}[n-l] + \mathbf{w}[n],
\end{equation}
where $\mathbf{x}[n]$ denotes the transmitted symbol vector across spatial streams, $\mathbf{H}[l]$ represents the channel impulse response at delay tap $l$, and $\mathbf{w}[n]$ denotes additive noise.

At the receiver, the incoming signal undergoes packet detection and matched filtering. Timing synchronisation and carrier frequency offset estimation are then performed using the training sequences in the preamble. Wireless channel is estimated using the known training symbols. Correlation between the received and transmitted sequences provides an estimate of the 
MIMO channel response $\hat{\mathbf{H}}[l] \in \mathbb{C}^{N_{Rx} \times N_{Tx}}, \quad l = 0, \dots, L-1,$ where $L$ is number of channel taps, $N_{Tx}$ and $N_{Rx}$ denote the numbers of transmit and receive antennas, respectively.

For frequency-selective channels, equalisation is performed in an appropriate transform domain to reduce computational complexity. In single-carrier systems, this is typically achieved via frequency-domain equalisation after a discrete Fourier transform (DFT), while in OFDM systems equalisation is naturally performed per subcarrier in the frequency domain, yielding the frequency-domain observation $\mathbf{y}[k]$, where $k$ denotes the frequency bin index. 
The estimated channel frequency response is given by $\hat{\mathbf{H}}[k] \in \mathbb{C}^{N_{Rx} \times N_{Tx}}$.

Typically, a minimum mean square error (MMSE) equaliser is employed, which uses an estimate of the noise variance $\sigma_n^2$ to compute the channel state information,
$\mathbf{CSI}[k] = \hat{\mathbf{H}}[k]\hat{\mathbf{H}}^{\mathrm{H}}[k] + \sigma_n^2 \mathbf{I},$
where $(\cdot)^{\mathrm{H}}$ denotes the Hermitian transpose and $\mathbf{I}$ is the identity matrix of appropriate dimension. The equalised frequency-domain symbols are obtained as
\begin{equation}
\hat{\mathbf{x}}[k] = \hat{\mathbf{H}}^{\mathrm{H}}[k]\mathbf{CSI}^{-1}[k]\mathbf{y}[k].
\end{equation}
The corresponding time-domain symbols are then obtained, either through an inverse discrete Fourier transform in single-carrier systems or directly from per-subcarrier equalisation in OFDM-based receivers, using $\hat{\mathbf{x}}[n] = \mathrm{IDFT}\{\hat{\mathbf{x}}[k]\}$.


After equalisation, a phase correction stage compensates for residual impairments such as phase noise or remaining carrier frequency offset. The corrected symbols are then passed to the demapping stage. 
The role of the demapper is to convert equalised complex symbols into soft bit metrics for channel decoding. In conventional receivers, this is commonly performed using approximate log-likelihood ratio (LLR) computation. The demapper operates on equalised symbols and is therefore independent of whether they originate from a single-carrier or OFDM-based receiver.


Among the receiver components, the demapping stage plays a key role, as it operates directly on the equalised symbols and is sensitive to residual distortions that persist after equalisation. Such distortions are difficult to capture with analytical models and can lead to inaccuracies in conventional demapping rules. Therefore, this work focuses on the design of the demapper, while all other receiver blocks are kept unchanged, preserving compatibility with conventional receiver architectures.

\subsection{Benchmark Demappers}\label{sec:benchmark}
This section reviews representative demapping approaches used in practical receivers, highlighting the limitations addressed in this work. These include the conventional 
soft demapper as well as neural network–based demappers that learn the mapping between received symbols and soft bit metrics from data.

\subsubsection{Conventional Soft Demapper}

The demapper converts the phase-corrected equalised complex symbols into soft bit metrics that are used by the channel decoder. Let $\tilde{x}[n]$ denote the phase-corrected equalised symbol obtained from $\hat{x}[n]$ for a given spatial stream at time index $n$. For an $M$-QAM constellation, each symbol represents $m=\log_2 M$ coded bits, and the demapper produces $m$ LLRs that quantify the reliability of each bit. 
%
In practical implementations, LLRs are approximated using the max-log approximation given 
by
\begin{equation}
LLR_i[n] \approx 
\max_{c \in \mathcal{C}_{i,0}}\!\left(-\frac{|\tilde{x}[n]-c|^2}{\sigma_n^2}\right)
-
\max_{c \in \mathcal{C}_{i,1}}\!\left(-\frac{|\tilde{x}[n]-c|^2}{\sigma_n^2}\right),
\end{equation}
where $\mathcal{C}_{i,0}$ and $\mathcal{C}_{i,1}$ denote the subsets of constellation points whose $i$th bit label equals 0 and 1, respectively, and $\sigma_n^2$ represents the noise variance.

\subsubsection{LLRNet Demapper}

LLRNet~\cite{llring} is a neural soft demapper that approximates the 
demapping operation using a compact feed-forward architecture. The input to the network is the equalised complex symbol $\tilde{x}[n]$, 
processed by a neural network consisting of a single hidden layer with $K$ neurons using a ReLU activation function, followed by a linear output layer with $m$ neurons that produce the LLRs corresponding to the transmitted bits.
The number of neurons in the hidden layer is selected according to the modulation order. 

\subsubsection{MLP-Based Demapper}

It is based on a multilayer perceptron (MLP) architecture~\cite{demapper2} that takes the real and imaginary components of the equalised symbol $\hat{x}[n]$ as input. The network is implemented as a feed-forward fully-connected architecture consisting of four hidden layers
, each employing a ReLU activation function. The output layer contains $m$ neurons and uses a linear activation to generate the LLRs.
Unlike LLRNet, the hidden-layer architecture of the MLP demapper remains fixed across modulation schemes. Only the size of the output layer changes with the modulation order
.


The above approaches show some key limitations. Symbol-wise demappers do not capture dependencies across neighbouring symbols, while joint-processing approaches introduce additional complexity. As a result, distortions in the received signal are not effectively exploited under practical conditions. To address this, Section \ref{sec:Proposed} introduces a block-wise demapper that exploits such dependencies while maintaining the modular structure of conventional receivers.

\section{Proposed DL-Based Demapper}\label{sec:Proposed}

In a conventional receiver, the demapper processes each equalised symbol independently and computes bit LLRs based solely on the current symbol and an estimate of the noise variance. This symbol-wise processing implicitly assumes statistical independence across symbols and relies on the additive white Gaussian noise (AWGN) model. While this assumption simplifies the demapping rule, it does not always reflect the behaviour of practical wideband systems, where receiver processing and hardware impairments can introduce distortions that deviate from the Gaussian noise assumption~\cite{hwImp_2}.

In practical receivers, several impairments remain even after synchronisation and equalisation. These include phase noise from local oscillators, nonlinear distortions introduced by power amplifiers, imperfect channel estimation, and residual inter-symbol interference. Such effects change the statistics of the received signal and create a mismatch between the assumed noise model used in LLR computation and the actual signal conditions. In addition, hardware-induced distortion noise typically scales with the signal power, which further deviates from the assumptions used in conventional demappers~\cite{hwImp_2, hwImp_emil}. 
As a result, the LLR values 
may become inaccurate, which can degrade the performance of the subsequent channel decoder. These limitations motivate data-driven demapping approaches that can adapt to the effective statistics of the received signal.

To address these limitations, we propose a deep learning–based demapper that operates on a block of equalised symbols, rather than processing symbols individually. The block can represent either a sequence of time-domain symbols in single-carrier systems or a set of subcarriers in multi-carrier (e.g., OFDM) systems. By jointly processing a block of symbols, the model can capture temporal dependencies introduced by residual channel effects, hardware impairments, and imperfect equalisation. This block-wise formulation enables the demapper to exploit contextual information across neighbouring symbols, producing more reliable soft bit estimates than symbol-wise demapping rules.


The proposed architecture is implemented using a one-dimensional convolutional neural network that processes blocks of equalised symbols. Convolutional layers extract local temporal features from the symbol sequence, enabling the network to capture residual distortions and channel memory effects. Compared with fully connected architectures, convolutional networks exploit local structure through shared filters, improving generalisation while reducing model complexity. By incorporating temporal context into the demapping process, the proposed demapper produces more reliable LLR estimates than conventional symbol-wise approaches.

\subsection{Block-wise MIMO Demapping Model}

We now describe the proposed block-wise demapping model. The formulation applies to both single-carrier and multi-carrier (e.g., OFDM) systems; for concreteness, we describe it using the single-carrier case. 
The demapper operates on blocks of phase-corrected equalised symbols collected from all spatial streams. Let $N_s$ denote the number of spatial streams, and let the phase-corrected equalised symbol vector at time index $n$ be denoted as
\begin{equation}
\begin{aligned}
\tilde{\mathbf{x}}[n] &=
\big[
\tilde{x}^{(1)}[n],
\tilde{x}^{(2)}[n],
\dots,
\tilde{x}^{(N_{s})}[n]
\big]^\top
\in \mathbb{C}^{N_{s}}, \\
&\quad n = 1,\dots,T,
\end{aligned}
\end{equation}
where $\tilde{x}^{(r)}[n]$ represents the phase-corrected equalised symbol corresponding to the $r$-th spatial stream at time index $n$, and $T$ denotes the number of symbols in a block.

Stacking these vectors over a block of $T$ symbols yields the complex-valued input matrix
\begin{equation}
\tilde{\mathbf{X}} =
\big[
\tilde{\mathbf{x}}_1,\dots,\tilde{\mathbf{x}}_T
\big]^\top
\in \mathbb{C}^{T \times N_{s}}.
\end{equation}

Each complex-valued entry of $\tilde{\mathbf{X}}$ is converted into a real-valued representation by separating its in-phase and quadrature components, and then concatenating along the feature dimension at each time index to form the real-valued tensor
\begin{equation}\label{eqn:demapIp}
\mathbf{X} \in \mathbb{R}^{T \times C_{\mathrm{in}}},
\end{equation}
which serves as the input to the proposed demapper, where $C_{\mathrm{in}} = 2N_{s}$ denotes the total number of input feature channels corresponding to the in-phase and quadrature components of all spatial streams.

The proposed demapper is parameterised by a neural network $f_{\boldsymbol{\theta}}$ that learns the mapping
\begin{equation}
    f_{\boldsymbol{\theta}} : \mathbb{R}^{T \times C_{\mathrm{in}}} \rightarrow \mathbb{R}^{T \times C_{out}},
\end{equation}
where $C_{out} = mN_{s}$ and $m$ is the number of bits per modulated symbol. The output consists of real-valued soft scores for each bit of each symbol in the block. This block-based mapping allows the model to exploit temporal structure that is inaccessible to symbol-wise demappers.

The demapper outputs soft information in the form of bit-wise probability estimates. These values serve as confidence measures for each bit and can be used either to derive hard decisions directly or, when required, to interface with a soft-input decoding stage~\cite{deeprx,llring}. As such, the proposed demapper can be integrated into a conventional receiver chain without structural changes, preserving the modularity of the receiver while introducing learning at a stage that directly affects detection performance.

This configuration, referred to as the multi-stream (MS) version, jointly processes all receive streams within a single model. While this allows the demapper to exploit potential cross-stream correlations, the input dimensionality scales with $N_{s}$. Consequently, the learned model is tied to a specific MIMO order and must be retrained or reconfigured if the number of receive antennas changes.


In addition to the MS configuration, we consider a single-stream (SS) version in which the demapper processes each receive stream independently. In this case, the input dimensionality is fixed to $C_{in}=2$, corresponding to the in-phase and quadrature components of a single stream, while the output dimensionality remains $C_{out}=m$. For a receiver with $N_{s}$ streams, the same demapper instance is applied independently to each stream. Since the architecture operates on a fixed two-channel input, it remains independent of the overall MIMO order and can therefore be directly applied to different antenna configurations without modification.

\subsection{Model Architecture}

The proposed demapper is implemented using a one-dimensional convolutional neural network with residual connections. The network operates on a fixed-length block of samples and preserves the temporal ordering of the input throughout the processing stages. All layers are convolutional, enabling parameter sharing across time and supporting an efficient hardware implementation.


The input to the model is the real-valued tensor $\mathbf{X}$ defined in Eqn.~\ref{eqn:demapIp}. The first stage consists of a one-dimensional convolution with kernel size one, followed by a ReLU activation:
\begin{equation}
\mathbf{z}_n = \phi(\mathbf{W}\mathbf{x}_n + \mathbf{b}),
\qquad \mathbf{W} \in \mathbb{R}^{D \times C{\mathrm{in}}},
\end{equation}
where $\mathbf{x}_n$ denotes the input feature vector at time index $n$, $\phi(\cdot)$ is the ReLU activation, and $D$ is the number of filters. This operation mixes information across feature channels while preserving the temporal structure of the symbol block.

The resulting feature sequence is processed by a stack of one-dimensional convolutional layers and a residual block. Convolutional layers with kernel size $S$ capture local temporal context across neighbouring symbols. 
The residual block consists of two convolution layers of kernel size $S$, where the first convolution is followed by a ReLU activation and a dropout layer, while the second convolution is applied without activation. A skip connection adds the block input to its output, allowing the network to learn residual feature refinements while preserving information across layers.


Following the residual block, an additional convolution layer further refines the feature representation. For hardware efficiency, all convolution layers except the input and output layers use a fixed number of filters, denoted by $N_{\text{Filt}}$. This uniform design enables reuse of the same convolution engine across layers, simplifying the hardware implementation.

The final layer is a one-dimensional convolution that maps the learned features to $C_{\mathrm{out}}$ output feature channels, producing $\mathbf{Z} \in \mathbb{R}^{T \times C_{\mathrm{out}}}$. Each element $Z_{n,k}$ corresponds to a soft score associated with bit $k$ of symbol $n$. 
The complete network structure is summarised in Table~\ref{tab:demapper_arch}.

\begin{table}[t]
\centering
\caption{Architecture of the proposed DL-based demapper. }
\label{tab:demapper_arch}
\resizebox{\linewidth}{!}{
\begin{tabular}{c c c c c}
\hline
Layer & Type & Kernel Size & Filters & Output Shape \\
\hline
Input  & --              & -- & --              & $T \times C_{in}$ \\
1      & Conv1D + ReLU   & 1  & $D$ & $T \times D$ \\
2      & Conv1D + ReLU   & S  & $N_{\text{Filt}}$ & $T \times N_{\text{Filt}}$ \\
3      & Residual Block  & S  & $N_{\text{Filt}}$ & $T \times N_{\text{Filt}}$ \\
4      & Conv1D + ReLU   & S  & $N_{\text{Filt}}$ & $T \times N_{\text{Filt}}$ \\
Output & Conv1D          & S  & $C_{out}$              & $T \times C_{out}$ \\
\hline
\addlinespace[0.5em]
\multicolumn{5}{l}{\textbf{Residual block internal structure:}} \\
\addlinespace[0.5em]
\hline
Step & Operation & Kernel Size & Filters & Notes \\
\hline
1 & Conv1D + ReLU & S & $N_{\text{Filt}}$ & -- \\
2 & Dropout      & -- & -- & rate = 0.1 \\
3 & Conv1D       & S & $N_{\text{Filt}}$ & -- \\
4 & Addition     & -- & -- & skip connection \\
\hline
\end{tabular}
}
\end{table}

\subsection{Training and Inference}

The demapper is trained in a supervised manner using the transmitted bits as ground-truth labels. The dataset is divided into disjoint training and test sets to ensure a strict separation between model training and evaluation. The training data are further split into training and validation subsets using an 80:20 ratio. Training is performed with a batch size of 64.


The dataset includes symbol blocks from all spatial streams of the MIMO receiver. In the MS configuration, symbols from all streams are jointly processed by a single demapper, enabling the model to learn cross-stream dependencies. In the SS configuration, each stream is processed independently, while data from all streams are combined within the training batch to enable parameter sharing across streams without requiring stream-specific models.

For each input block $\mathbf{X}$, the demapper produces a tensor of real-valued outputs $\mathbf{Z} \in \mathbb{R}^{T \times C_{\mathrm{out}}}$, where each element represents a soft score associated with a bit of a symbol in the block. During training, these logits are passed through a sigmoid function to obtain soft bit probabilities. The network parameters $\boldsymbol{\theta}$ are optimised using the binary cross-entropy loss in its logits form, which improves numerical stability and is standard in learning-based detection.

At inference time, the sigmoid operation is omitted and the logits $\mathbf{Z}$ are used directly. These values can be interpreted as soft bit scores and passed to a soft-input LDPC decoder, or converted to hard decisions when required.

Model optimisation is performed using the Adam optimiser with an initial learning rate of $2\times10^{-3}$. A learning-rate scheduler reduces the learning rate by a factor of five when the validation loss plateaus, with a patience of four epochs. Early stopping monitors the validation loss with a patience of 15 epochs to prevent overfitting. The model corresponding to the lowest validation loss is retained and used for evaluations.

\subsection{Hardware Implementation and Hardware–Software Co-Design}\label{sec:hwDesign}
To demonstrate the practical feasibility of the proposed deep learning–based demapper, the inference stage of the model is implemented as hardware accelerator on the MIMORPH platform~\cite{mimorph}. The design focuses on realising the DL-based demapper efficiently while maintaining low hardware complexity and compatibility with the receiver pipeline.

The accelerator architecture is built around a single configurable one-dimensional convolution engine that is reused sequentially across all layers. This design avoids dedicated compute units for each layer and therefore significantly reduces hardware requirements. Although the first convolution layer uses a kernel size of one, it is mapped onto the same convolution engine by padding unused kernel coefficients with zeros, enabling all layers to share a uniform datapath.

Convolution weights and biases are stored in on-chip block RAM (BRAM) and accessed using layer-specific offsets during execution. Intermediate feature maps are stored in shared on-chip buffers operating in a ping–pong configuration. This buffering scheme minimises on-chip memory usage while enabling continuous processing of feature maps. Input feature blocks are transferred from external memory only once at the beginning of processing, while intermediate activations remain on-chip until the final outputs are produced. The accelerator is implemented as a reusable hardware IP core exposing AXI-Stream interfaces, allowing it to operate as a streaming processing block within an FPGA-based receiver pipeline.

\begin{figure}[]
    \centering
    \includegraphics[width=0.9\linewidth]{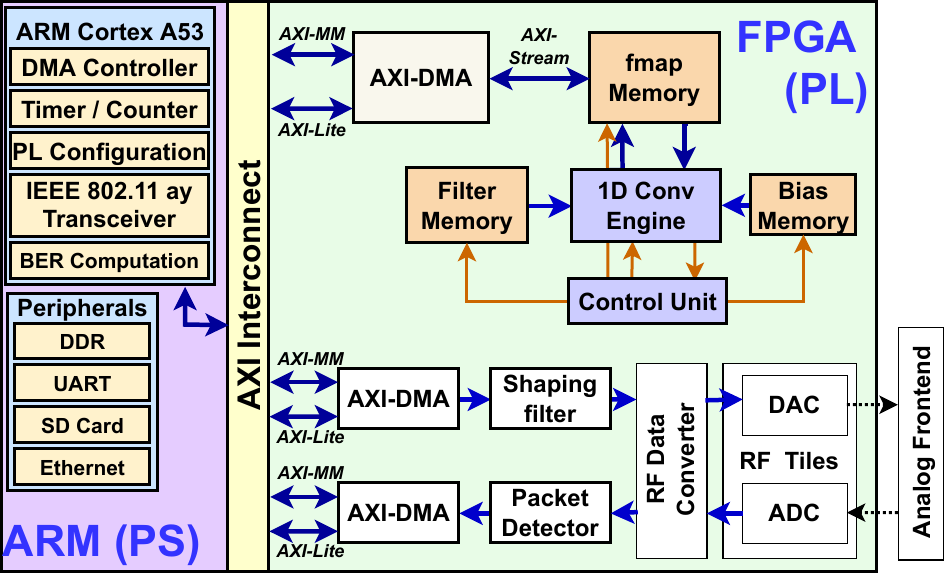}
    \label{fig:hscd}
    \caption{\small System-level PS–PL hardware architecture with the proposed DL-based demapper implemented in the PL.}
\end{figure}

The accelerator is implemented on an RFSoC, which integrates a processing system (PS) with programmable logic (PL), enabling a hardware–software co-design architecture as illustrated in Fig.~\ref{fig:hscd}. In the adopted system configuration, the PS executes the conventional receiver processing chain, including packet detection, synchronisation, channel estimation, equalisation, and phase correction. The resulting phase-corrected symbol blocks are stored in external DDR memory and streamed to the PL-based demapper accelerator using an AXI Direct Memory Access (DMA) engine. After processing, the soft outputs generated by the accelerator are written back to DDR memory through the same DMA mechanism, from where subsequent receiver processing continues on the PS.

While the adopted PS–PL partitioning introduces additional data movement between PS memory and the PL, it isolates the proposed demapper accelerator from the remaining receiver pipeline, enabling independent evaluation of its computational cost and hardware efficiency. In a fully integrated implementation, additional baseband processing stages could be migrated to the PL to reduce memory transfers and improve end-to-end latency. However, the focus of this work is to evaluate the proposed demapper architecture as an independent accelerator.

From a system perspective, the hardware accelerator operates as a drop-in replacement for conventional demappers. It preserves the same block-wise input–output interface, supports different modulation orders
, and remains independent of the number of receive antennas, since the same trained model is applied per spatial stream. The accelerator exposes standard AXI interfaces, enabling seamless integration into FPGA-based receiver pipelines as a replacement for conventional demappers~\cite{mobisys_helix}, as well as deployment as a hardware accelerator in software-centric receiver architectures such as those used in disaggregated and O-RAN–style systems. Detailed micro-architectural aspects, including multiply–accumulate scheduling and register-level optimisations, are omitted for brevity, as the focus of this section is on the system-level architecture and integration of the proposed demapper.

\section{Experimental Setup and Over-the-Air Dataset}\label{sec:testbed}
The proposed demapper is evaluated using an IEEE 802.11ay~\cite{802_11ay} single-carrier mmWave MIMO waveform as a representative wideband communication model. The system operates at a carrier frequency of 60~GHz with a $4\times4$ MIMO configuration and four spatial streams ($N_s = N_{Rx} = 4$), where each spatial stream occupies 1.76~GHz of bandwidth.

Each transmit packet consists of a preamble for synchronisation and channel estimation, followed by a payload carrying the encoded data symbols. The payload is organised into blocks of 512 symbols, comprising 64 guard symbols and 448 data symbols. For 16QAM transmission, each packet contains 8 blocks, corresponding to 14,336 transmitted bits, while for 64QAM each packet contains 5 blocks, corresponding to 13,440 transmitted bits.
Channel estimation is performed in the time domain using orthogonal Golay sequences of length 128, producing a MIMO channel impulse response of size $4\times4\times128$. 

\subsection {Measurement Testbed}

\begin{figure}
    \centering
    \includegraphics[width=\linewidth]{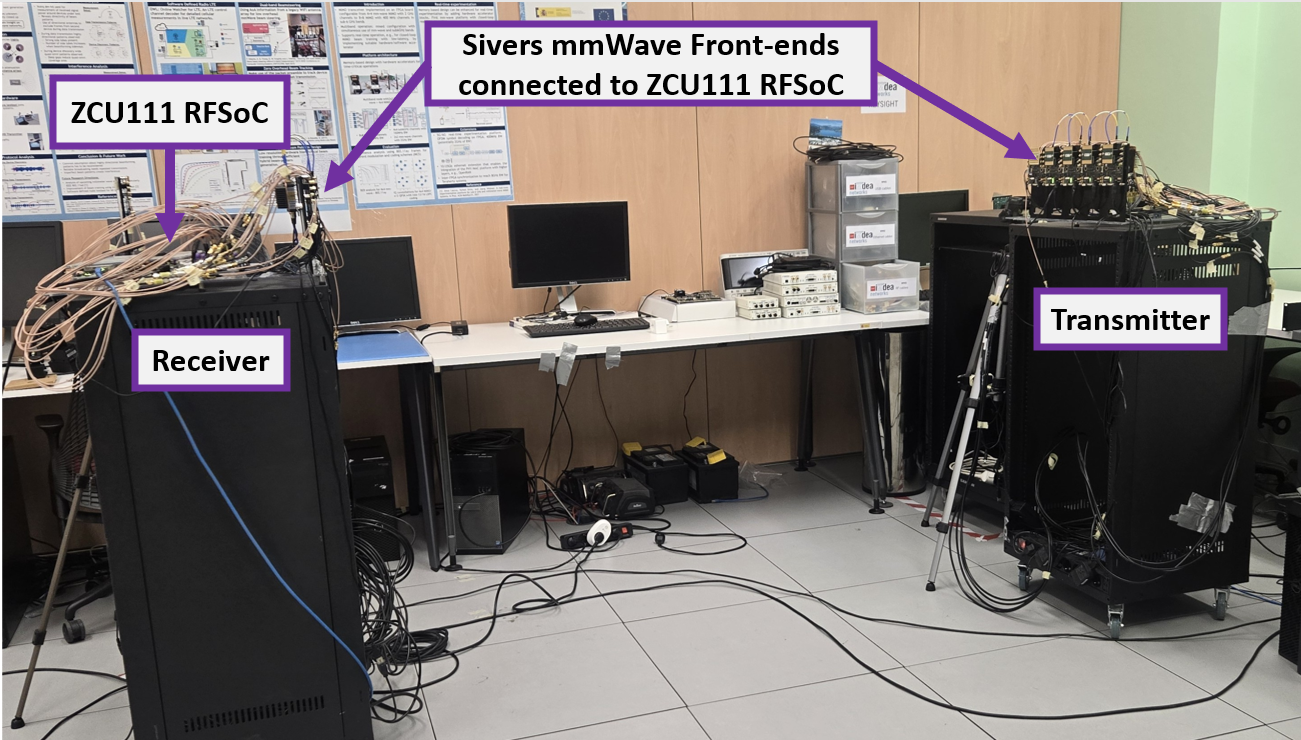}
    \caption{\small Experimental setup used to collect the training and validation OTA dataset, deployed in an indoor laboratory environment.}
    \label{fig:lab}
\end{figure}

All measurements are conducted using the in-house MIMORPH~\cite{mimorph} millimetre-wave testbed, an open-source FPGA-based platform for wideband mmWave MIMO experimentation. The testbed is built around an AMD Xilinx RFSoC and uses Sivers Semiconductors 60~GHz RF front ends at both the transmitter and receiver. 
The testbed operates in a memory-based configuration. At the transmitter, baseband packets are pre-generated and stored in on-board memory before being streamed through the RF chain. At the receiver, packet detection is performed in FPGA logic, after which the corresponding complex in-phase and quadrature samples are captured and stored for offline processing. This configuration preserves access to minimally processed received signals.

Measurements are conducted across multiple indoor and outdoor environments with a clear line of sight between the transmitter and receiver, and across multiple days and transmit gain settings. These variations introduce diversity in propagation conditions and hardware states, enabling evaluation under practical operating scenarios. The laboratory setup used to collect the dataset is shown in Fig.~\ref{fig:lab}.

\subsection{Dataset Generation}

The dataset is generated from baseband packets captured by the MIMORPH receiver. After packet detection in the FPGA, the corresponding complex in-phase and quadrature samples are stored and processed offline. Each captured packet is then processed using the receiver chain described in Section~\ref{sec:sysDesign} to obtain the equalised symbol sequences. The corresponding transmitted bit sequences are recovered from the stored transmit packets, enabling the construction of labelled input–output pairs. For each payload block, the dataset stores (i) the equalised complex symbols, (ii) the associated channel frequency response, and (iii) the transmitted uncoded bit sequence. These elements form the input features and labels used for supervised training and evaluation of the demapper.

\section{Results}\label{sec:results}

This section presents the performance evaluation of the proposed deep learning–based demapper using over-the-air measurements collected with the $4\times4$ MIMO mmWave system defined in Section~\ref{sec:testbed}. All results are obtained on a held-out test set collected on different days from the training data, while using the same experimental setup. This separation prevents data leakage and ensures that evaluation reflects channel conditions, hardware states, and noise characteristics not observed during training. Performance is reported in terms of bit error rate (BER) measured after the complete receiver chain, including equalisation, phase correction, demapping, and decoding. BER is computed on a per-frame basis and aggregated across all spatial streams for analysis and plotting.
To quantify the performance improvement of the proposed scheme over the conventional baseline, we define the BER Gain, denoted as $\mathcal{G}_{BER}$, as the ratio of the baseline BER to the evaluated scheme's BER:
\begin{equation}
    \mathcal{G}_{BER} = \frac{BER_{baseline}}{BER_{eval}}
\end{equation}

For performance visualisation, packets are organised according to an estimated signal-to-interference-plus-noise ratio (SINR). This SINR is computed only for analysis and is not provided as input to the demapper. Let $x[n]$ denote the transmitted constellation symbol and $\tilde{x}[n]$ the corresponding received symbol after equalisation and phase correction, for $n = 1, \dots, T$. The error power is defined as
\begin{equation}
    P_e = \frac{1}{T} \sum_{n=1}^{T} | \tilde{x}[n] - x[n] |^2,
\end{equation}
and the signal power is
\begin{equation}
    P_s = \frac{1}{T} \sum_{n=1}^{T} | x[n] |^2.
\end{equation}
The estimated SINR is then given by
\begin{equation}
    \text{SINR} = 10 \log_{10} \left( \frac{P_s}{P_e} \right).
\end{equation}
Using the known transmitted symbols to compute SINR is practical for offline evaluation and does not influence the learned model or the receiver operation.
To construct BER curves, packets are sorted by their estimated SINR and grouped into bins. For each bin, the BER is computed as the ratio of the total number of erroneous bits to the total number of transmitted bits across all packets within that bin.

\subsection{Performance Evaluation}

\begin{figure}[t]
    \centering
    \includegraphics[width=0.85\linewidth]{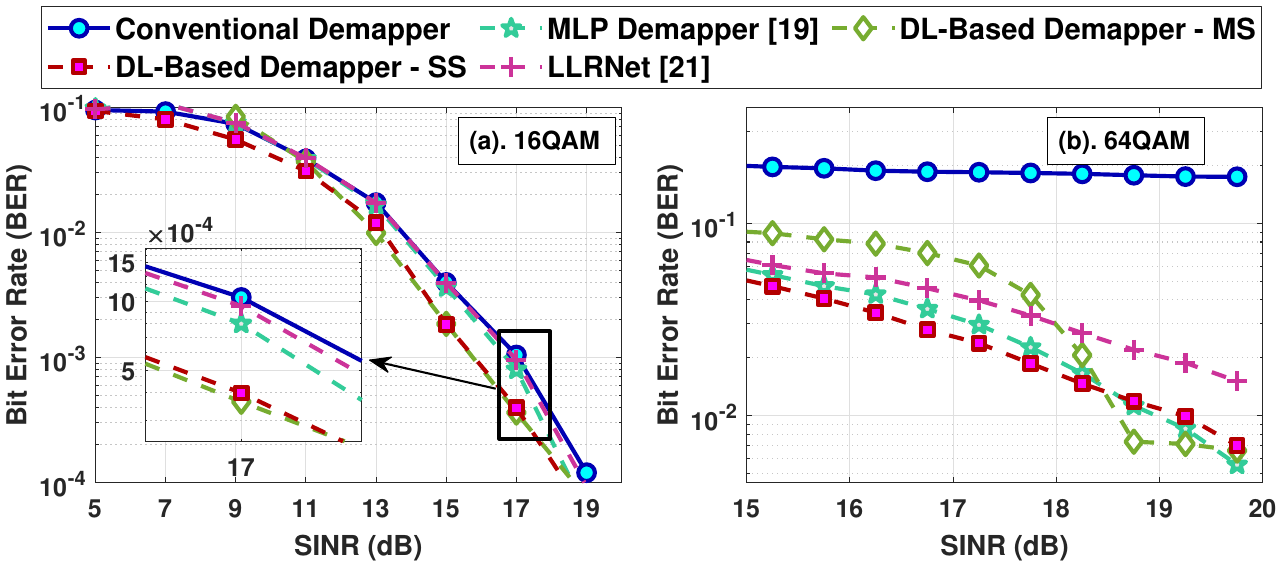}
    \vspace{-2mm}
    \caption{\small BER versus estimated SINR for the conventional demapper and the proposed DL-based demapper. (a) 16QAM, (b) 64QAM.}
    \label{fig:perfAll}
\end{figure}

We present the performance evaluation of the DL-based demapper under practical operating conditions using OTA data. Fig.~\ref{fig:perfAll} shows the BER as a function of SINR for the proposed single-stream and multi-stream variants of the DL-based demapper, along with MLP-based demappers~\cite{demapper2,llring}. The MLP-based demappers operate on a symbol-by-symbol basis without exploiting any block level context. 

We compare the single-stream and multi-stream variants of the proposed DL-based demapper. For both the single-stream and multi-stream variants, $N_{Filt} = 64$. For 16QAM, both variants exhibit very similar performance across the entire SINR range. The single-stream variant achieves a maximum BER gain of 2.6$\times$ and an average gain of 1.8$\times$ over the conventional demapper, while the multi-stream variant achieves a maximum gain of 2.9$\times$ and an average gain of 1.7$\times$.

For 64QAM, clear differences emerge between the two variants. The single-stream version shows better performance over most of the SINR range, with a maximum BER gain of 2.5$\times$ over the multi-stream variant. The BER of the multi-stream variant decreases rapidly around 18 dB SINR but starts flattening around 19 dB, whereas the single-stream exhibits a steady reduction in BER across the entire SINR range. {The flattening of the multi-stream demapper can be attributed to the fact that, after MIMO equalisation, the residual distortions affecting each spatial stream are largely \textit{stream-local}. While the multi-stream architecture enables joint processing across streams, the additional inputs introduce variability that is not consistently informative for per-stream bit decisions. At high SINR, where noise is no longer dominant, this increased input dimensionality can lead to small systematic mismatches in the learned decision regions, particularly for higher-order constellations}. Compared to the conventional demapper, the multi-stream variant achieves a maximum BER gain of 24.5$\times$ and an average gain of 7.7$\times$, while the single-stream variant achieves a maximum gain of 17.6$\times$ and an average gain of 8.7$\times$.

These observations indicate that jointly processing multiple receive streams does not provide a consistent advantage for demapping in the considered system. Cross-stream interference is largely mitigated by the preceding channel estimation and equalisation stages, resulting in streams that are approximately independent at the demapper input. Concatenating all receive streams therefore increases the input dimensionality and learning complexity without introducing additional exploitable structure.

We now compare the proposed DL-based demapper with two neural network-based demappers\cite{demapper2,llring} described in section~\ref{sec:benchmark}. For 16QAM, 
the single-stream and multi-stream DL variants achieve maximum BER gains of 2.0$\times$ and 2.2$\times$, respectively, over the MLP-based demapper~\cite{demapper2} with corresponding average gains of 1.5$\times$ and 1.4$\times$.
When compared with LLRNet~\cite{llring}, the proposed DL-based demapper achieves maximum BER gains of 2.4$\times$ and 2.6$\times$ for the single and multi-stream variants, respectively, with corresponding average gains of 1.7$\times$ and 1.6$\times$.
For 64QAM, the improvements of the proposed DL-based demapper over the MLP model are more moderate. The single and multi-stream variants achieve maximum BER gains of 1.3$\times$ and 1.5$\times$, respectively, with average gains of 1.1$\times$ and 0.75$\times$ compared to the MLP-based demapper~\cite{demapper2}.
In comparison with LLRNet~\cite{llring}, the single-stream and multi-stream variants achieve maximum BER gains of 2.1$\times$ and 3.4$\times$, respectively, with average gains of 1.7$\times$ and 1.3$\times$.
These results show that while symbol wise learning can provide certain gains, especially at higher modulation orders, block wise processing enables more robust and consistent performance improvements under realistic OTA channel conditions.

In addition to these performance trends, it is also important to consider the practical implications of the proposed architectures.
From a practical perspective, the single-stream approach remains more attractive. Its complexity and parameter count do not scale with the number of receive antennas, since the same trained model is reused across streams. This makes the single-stream demapper scalable to different MIMO orders and well suited for resource constrained implementations, while maintaining strong performance in OTA measurements. Therefore, in the remainder of this paper, the single-stream version is adopted unless stated otherwise.

\begin{figure}[t]
    \centering
    \includegraphics[width=0.85\linewidth]{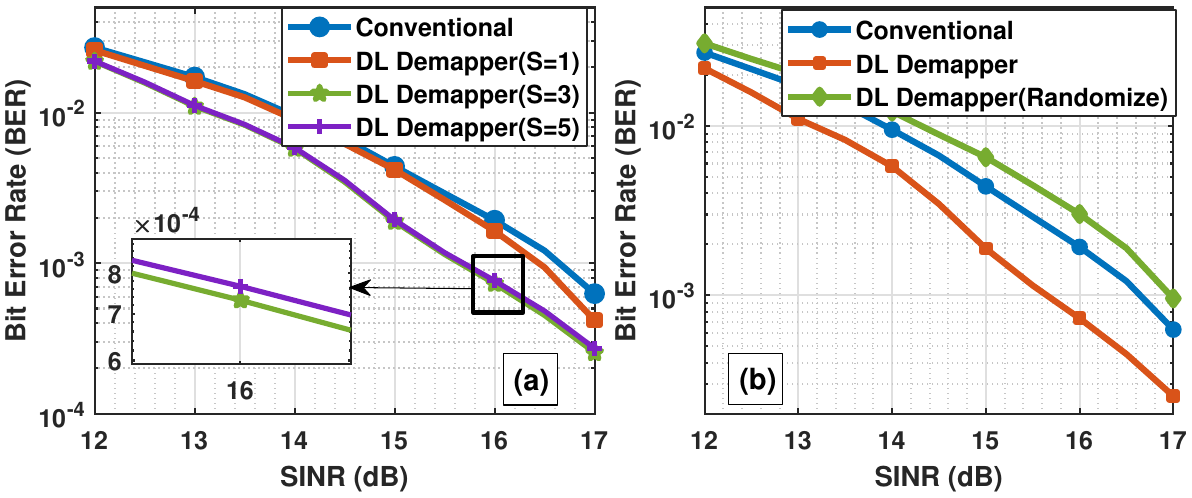}
    \vspace{-2mm}
    \caption{\small BER comparison between symbol-wise and block-wise processing using the DL-based demapper for 16QAM.}
    \label{fig:16qam_k1}
\end{figure}

To isolate the impact of block-wise processing, we evaluate the effect of the convolution kernel size on the BER performance of the proposed demapper. Figure~\ref{fig:16qam_k1}(a) shows results for kernel sizes $S=1$, $S=3$, and $S=5$, which determine the receptive field of the network. When $S=1$, the receptive field is limited to a single symbol and the network processes each symbol independently. In this case, the BER of the DL-based demapper is nearly identical to that of the conventional soft-decision demapper across most of the SINR range, with only a slight improvement at high SINR. In this regime the channel conditions are relatively clean and the dominant distortions arise from hardware imperfections and residual receiver effects, which the data-driven model can partially learn even without temporal context.
Increasing the kernel size to $S=3$ expands the receptive field so that each symbol is processed jointly with neighbouring symbols in the block. In this configuration the DL-based demapper consistently outperforms the conventional demapper across the entire SINR range, indicating that the model exploits temporal correlations in the received data to produce more reliable LLR estimates. Further increasing the kernel size to $S=5$ (receptive field $=21$) provides almost identical performance to $S=3$, suggesting that the relevant temporal dependencies are captured within a relatively small neighbourhood of symbols.

To verify that the performance gain arises from temporal information, we randomise the order of symbols within each block before providing them to the demapper. As shown in Fig.~\ref{fig:16qam_k1}(b), this operation has no effect on the conventional demapper since it processes symbols independently. In contrast, the BER of the DL-based demapper degrades and falls below the conventional demapper, confirming that the proposed model relies on temporal relationships between neighbouring symbols to improve LLR estimation.

\begin{figure*}[t]
    \centering
    \includegraphics[width=0.85\linewidth]{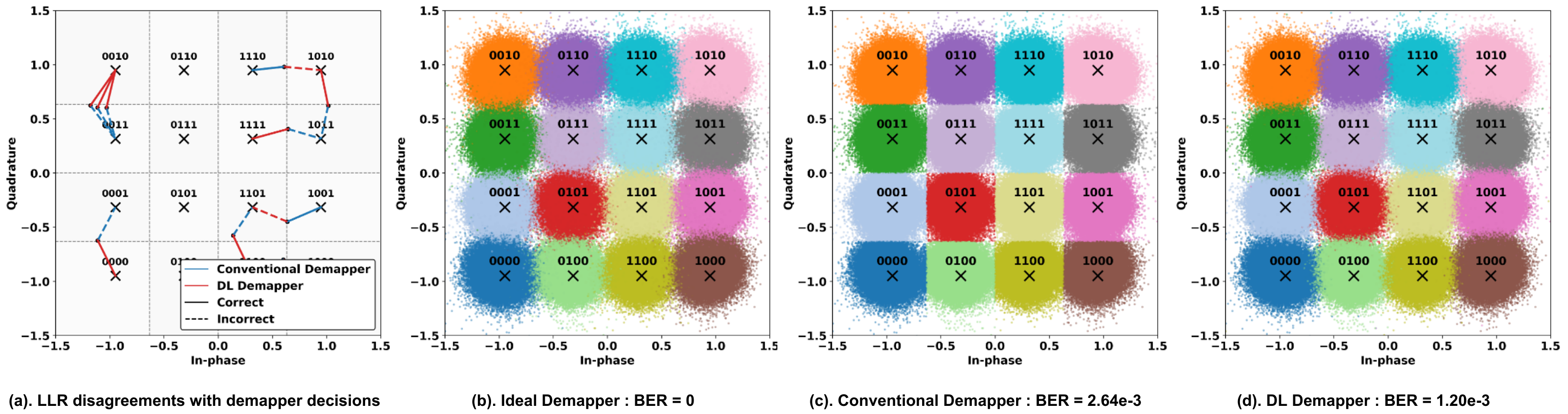}
    \caption{\small (a) Constellation points where the two demappers produce different bit decisions, (b) Ideal constellation labelled by true transmitted bits, (c) decisions of the conventional demapper, and (d) decisions of the DL-based demapper.}
    \label{fig:constellation}
\end{figure*}

To gain further insight into the decision behaviour of the two demappers, we analyse the constellation regions where their bit decisions differ. For each payload, the phase-corrected symbols are recorded and the corresponding bit decisions from both demappers are compared. Figure~\ref{fig:constellation}(a) shows a subset of the constellation points where the two demappers produce different decisions. Most disagreements occur near the decision boundaries, where residual noise and distortion make symbol interpretation ambiguous. In these regions, both demappers can occasionally make incorrect decisions; however, the DL-based demapper is correct in most cases, as also evident from several examples highlighted in the figure. While there are instances where the conventional demapper produces the correct decision, these are less frequent, indicating that the DL-based approach is better able to resolve ambiguous symbols and provide more reliable decisions overall.

To better understand these differences, Fig.~\ref{fig:constellation}(b),(c),(d) compares the received constellation diagrams under different demapping strategies. In Fig.~\ref{fig:constellation}(b), each received symbol is coloured according to its true transmitted label, representing the ideal decision regions in the presence of channel distortions and hardware impairments, which appear irregular. In Fig.~\ref{fig:constellation}(c), the conventional demapper applies fixed analytical decision boundaries, resulting in regular and symmetric regions that do not align well with the distorted constellation. In contrast, Fig.~\ref{fig:constellation}(d) shows that the DL-based demapper produces irregular decision regions that more closely follow the ideal structure in (b), indicating that it adapts its decision boundaries to the actual distribution of the received symbols. This closer alignment with the ideal regions explains the improved BER performance, as the DL-based demapper is better able to handle distortions and ambiguous symbols than the conventional approach.



\begin{figure}[t]
    \centering
    \includegraphics[width=0.95\linewidth]{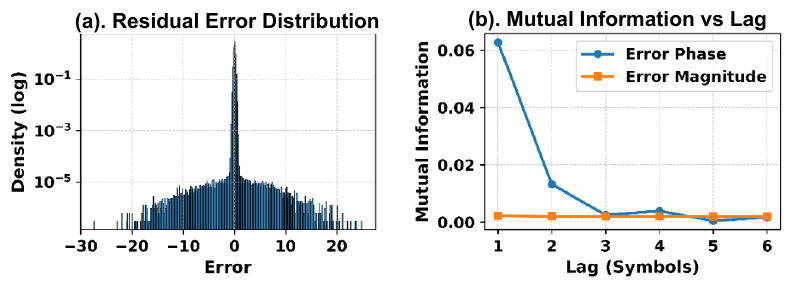}
    \vspace{-2mm}
    \caption{\small Residual error and its impact on demapping performance: (a)error distribution, and (b) mutual information across symbol lags.}
    \vspace{-2mm}
    \label{fig:ResError}
\end{figure}

To further understand the observed performance gains, we analyze the residual error characteristics and their temporal structure. The log-scale error distribution in Fig.~\ref{fig:ResError}(a) shows that, although most residual errors remain small, a fraction of large deviations exists, indicating non-Gaussian behavior. To investigate whether these residual distortions exhibit exploitable structure, Fig.~\ref{fig:ResError}(b) shows the mutual information across symbol lags. The results reveal negligible dependency in the error magnitude but a weak short-range dependency in the error phase, particularly at lag one. This suggests that the proposed block-wise demapper benefits from exploiting localized structure in the residual error through joint processing of neighbouring symbols.

\subsection{Side Information addition}\label{sec:results_sideChannel}

We next investigate whether the proposed demapper can benefit from additional side information beyond the equalised payload symbols. Two forms of side information are considered. First the channel frequency response (CFR), which captures per-symbol channel variations across transmit antennas. The second is the guard interval (GI), which experiences the same channel and hardware impairments as the payload. Residual distortions observed on the received GI after equalisation and phase correction may provide information about impairments affecting the data symbols.

To incorporate the GI, the real and imaginary components of both the transmitted and received guard sequences are concatenated to form a 256-dimensional vector per block. This vector is passed through a learnable linear projection to match the payload length and the projected output is appended as an additional input channel to the demapper. The projection weights are learned jointly with the demapper parameters during training.

The CFR is incorporated by separating the real and imaginary parts of the channel coefficients for each transmit stream and appending them along the feature channel dimension. This increases the input dimensionality from  $C_{in}$ to $C_{in} + 2N_{s}$. 
In the $4\times4$ MIMO configuration with 4 spatial streams, this results in an increase of $C_{in}$ from 2 to 10, because of the real and imaginary parts of the channel coefficients. The resulting complexity overhead is small. For both 16QAM ($m$=4) and 64QAM ($m$=6), the CFR-assisted demapper increases the number of parameters and FLOPs by approximately 1\% compared to the baseline model.
In contrast, the GI-assisted configuration leads to a much larger increase in model size. The number of trainable parameters more than double, relative to the baseline, while the FLOP count increases by less than 1\%. These results are summarised in Table~\ref{tab:sideinfo_complexity}.

\begin{table}[b]
\vspace{-15pt}
\caption{\small Model complexity for different side-information configurations.}
\vspace{-5pt}
\renewcommand{\arraystretch}{1.2}
\label{tab:sideinfo_complexity}
 \resizebox{\columnwidth}{!}{
\begin{tabular}{|l|ll|ll|}
\hline
\multirow{2}{*}{\textbf{Side Info}} & \multicolumn{2}{c|}{\textbf{M=4}} & \multicolumn{2}{c|}{\textbf{M=6}} \\ \cline{2-5} 
 & \multicolumn{1}{l|}{\textbf{Parameters}} & \textbf{FLOPs} & \multicolumn{1}{l|}{\textbf{Parameters}} & \textbf{FLOPs} \\ \hline
\textbf{None (Baseline)} & \multicolumn{1}{l|}{50372} & 44986368 & \multicolumn{1}{l|}{50758} & 45330432 \\ \hline
\textbf{CFR} & \multicolumn{1}{l|}{50884} & 45445120 & \multicolumn{1}{l|}{51270} & 45789184 \\ \hline
\textbf{GI} & \multicolumn{1}{l|}{165572} & 45273088 & \multicolumn{1}{l|}{165958} & 45617152 \\ \hline
\textbf{CFR increase} & \multicolumn{1}{l|}{1.02\%} & 1.02\% & \multicolumn{1}{l|}{1.01\%} & 1.01\% \\ \hline
\textbf{GI increase} & \multicolumn{1}{l|}{228.70\%} & 0.64\% & \multicolumn{1}{l|}{226.96\%} & 0.63\% \\ \hline
\end{tabular}
}
\end{table}

\begin{figure}[]
    \centering
    \includegraphics[width=0.9\linewidth]{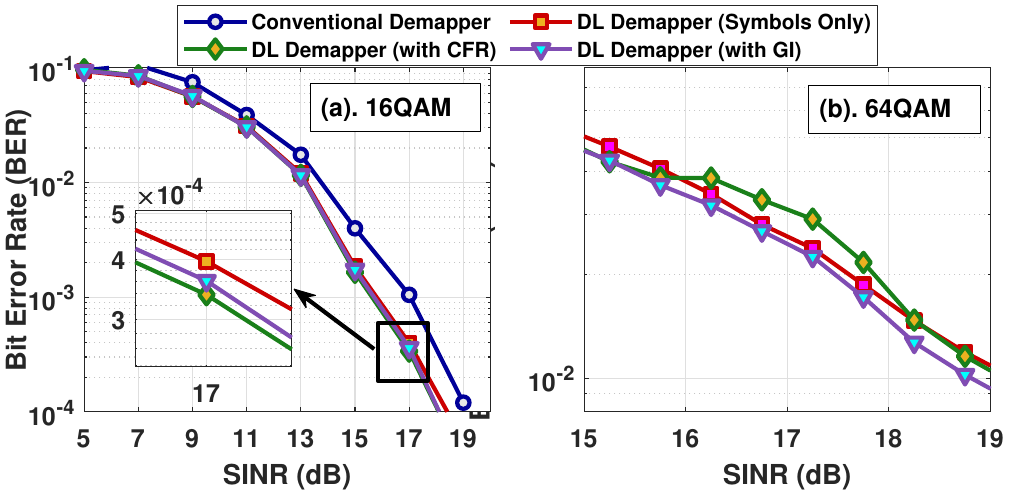}
    \caption{\small Effect of side information on performance of DL-Demapper.  Modulation : (a) 16QAM; (b) 64QAM.}
    \vspace{-2mm}
    \label{fig:sideInfo}
\end{figure}

The resulting BER performance is shown in Fig.~\ref{fig:sideInfo}. For 16QAM, Fig.~\ref{fig:sideInfo}(a) shows that incorporating the GI yields a small but consistent BER reduction across the SINR range, with the addition of CFR providing a marginal further improvement. However, the overall gain remains limited. A similar behaviour is observed for 64QAM in Fig.~\ref{fig:sideInfo}(b). The GI-assisted demapper provides a BER reduction of 1.2$\times$, while the CFR-assisted configuration performs nearly identically to the baseline model and, in some SINR regions, slightly worse.

These results indicate that the proposed block-wise demapper already extracts most of the relevant information directly from the received payload samples. As a result, explicitly providing side information offers only marginal benefit and does not yield a favourable trade-off between performance improvement and model complexity. This observation supports the design choice of relying primarily on the received waveform itself, which simplifies the receiver and avoids the need for additional inputs that may not be available or reliable in practical deployments.

\subsection{Ablation Study} \label{sec:ablation}

\begin{figure}[b]
    \centering
    \includegraphics[width=0.9\linewidth]{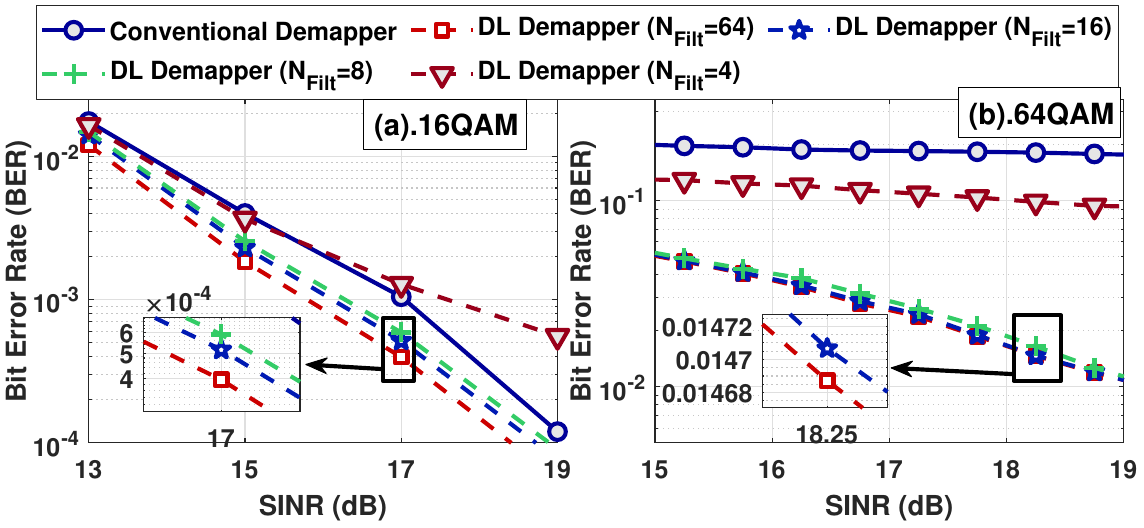}
    \caption{\small BER versus SINR performance for the proposed DL-based demapper with $N_{Filt}={64,16,8,4}$ for (a) 16QAM and (b) 64QAM, illustrating the impact of model size on detection performance.
}
\vspace{-2mm}
    \label{fig:ablation}
\end{figure}

\begin{table*}[]
\caption{Complexity and BER gain comparison for different $N_{Filt}$ values in the proposed DL-based demapper, relative to the $N_{Filt}=64$ baseline.}
\renewcommand{\arraystretch}{1.3}
\setlength{\tabcolsep}{3pt}
\label{tab:ablation}
 \resizebox{\textwidth}{!}{
\begin{tabular}{|l|ccccccc|cccclcc|}
\hline
\multirow{2}{*}{\textbf{$N_{Filt}$}} & \multicolumn{7}{c|}{\textbf{16QAM}} & \multicolumn{7}{c|}{\textbf{64QAM}} \\ \cline{2-15} 
 & \multicolumn{1}{c|}{\textbf{Params}} & \multicolumn{1}{c|}{\textbf{FLOPs}} & \multicolumn{1}{c|}{\textbf{\begin{tabular}[c]{@{}c@{}}Params\\ Reduction\end{tabular}}} & \multicolumn{1}{c|}{\textbf{\begin{tabular}[c]{@{}c@{}}FLOPs\\ Reduction\end{tabular}}} & \multicolumn{1}{c|}{\textbf{\begin{tabular}[c]{@{}c@{}}Min BER\\ Gain\end{tabular}}} & \multicolumn{1}{c|}{\textbf{\begin{tabular}[c]{@{}c@{}}Max BER\\ Gain\end{tabular}}} & \textbf{\begin{tabular}[c]{@{}c@{}}Av. BER\\ Gain\end{tabular}} & \multicolumn{1}{c|}{\textbf{Params}} & \multicolumn{1}{c|}{\textbf{FLOPs}} & \multicolumn{1}{c|}{\textbf{\begin{tabular}[c]{@{}c@{}}Params\\ Reduction\end{tabular}}} & \multicolumn{1}{c|}{\textbf{\begin{tabular}[c]{@{}c@{}}FLOPs\\ Reduction\end{tabular}}} & \multicolumn{1}{c|}{\textbf{\begin{tabular}[c]{@{}c@{}}Min BER\\ Gain\end{tabular}}} & \multicolumn{1}{c|}{\textbf{\begin{tabular}[c]{@{}c@{}}Max BER\\ Gain\end{tabular}}} & \textbf{\begin{tabular}[c]{@{}c@{}}Av. BER\\ Gain\end{tabular}} \\ \hline
\textbf{64} & \multicolumn{1}{c|}{50372} & \multicolumn{1}{c|}{45.0M} & \multicolumn{1}{c|}{-} & \multicolumn{1}{c|}{-} & \multicolumn{1}{c|}{1.018} & \multicolumn{1}{c|}{2.638} & 1.660 & \multicolumn{1}{c|}{50758} & \multicolumn{1}{c|}{45.3M} & \multicolumn{1}{c|}{-} & \multicolumn{1}{c|}{-} & \multicolumn{1}{l|}{3.723} & \multicolumn{1}{c|}{17.583} & 8.710 \\ \hline
\textbf{16} & \multicolumn{1}{c|}{\textbf{3380}} & \multicolumn{1}{c|}{\textbf{3.0M}} & \multicolumn{1}{c|}{\textbf{93.30\%}} & \multicolumn{1}{c|}{\textbf{93.40\%}} & \multicolumn{1}{c|}{\textbf{1.016}} & \multicolumn{1}{c|}{\textbf{2.029}} & \textbf{1.364} & \multicolumn{1}{c|}{\textbf{3478}} & \multicolumn{1}{c|}{\textbf{3.1M}} & \multicolumn{1}{c|}{\textbf{93.20\%}} & \multicolumn{1}{c|}{\textbf{93.20\%}} & \multicolumn{1}{l|}{\textbf{3.650}} & \multicolumn{1}{c|}{\textbf{17.837}} & \textbf{8.634} \\ \hline
\textbf{8} & \multicolumn{1}{c|}{924} & \multicolumn{1}{c|}{0.8M} & \multicolumn{1}{c|}{98.20\%} & \multicolumn{1}{c|}{98.20\%} & \multicolumn{1}{c|}{1.006} & \multicolumn{1}{c|}{1.776} & 1.258 & \multicolumn{1}{c|}{974} & \multicolumn{1}{c|}{0.8M} & \multicolumn{1}{c|}{98.10\%} & \multicolumn{1}{c|}{98.10\%} & \multicolumn{1}{l|}{3.590} & \multicolumn{1}{c|}{17.185} & 8.064 \\ \hline
\textbf{4} & \multicolumn{1}{c|}{272} & \multicolumn{1}{c|}{0.2M} & \multicolumn{1}{c|}{99.50\%} & \multicolumn{1}{c|}{99.50\%} & \multicolumn{1}{c|}{0.216} & \multicolumn{1}{c|}{1.092} & 0.916 & \multicolumn{1}{c|}{298} & \multicolumn{1}{c|}{0.3M} & \multicolumn{1}{c|}{99.40\%} & \multicolumn{1}{c|}{99.40\%} & \multicolumn{1}{l|}{1.527} & \multicolumn{1}{c|}{1.900} & 1.692 \\ \hline
\end{tabular}
}
\end{table*}

We examine the impact of model size by varying the number of filters, $N_{Filt}$, used in the convolution blocks of the proposed demapper. The baseline used in previous sections employs $N_{Filt}=64$. To assess the sensitivity to network width, we evaluate reduced configurations with $N_{Filt} \in \{16,8,4\}$ while keeping all other architectural components and training settings unchanged. The resulting performance and complexity trade-offs are summarised in Table~\ref{tab:ablation}.


For 16QAM, the configuration with $N_{Filt}=16$ maintains BER performance close to the baseline, with the minimum BER gain remaining above unity. The average BER gain decreases from 1.660 for $N_{Filt}=64$ to 1.364 for $N_{Filt}=16$, while the maximum BER gain also reduces moderately. When reduced further to $N_{Filt}=8$, the average BER gain drops to 1.258 and the minimum BER gain approaches unity, indicating reduced robustness. For $N_{Filt}=4$, both the minimum and average BER gains fall below unity, showing that the model no longer consistently outperforms the conventional demapper.


For 64QAM, configurations with $N_{Filt}=\{64,16,8\}$ exhibit similar performance, with minimum BER gains remaining well above unity and average BER gains varying only slightly. In particular, the average BER gain changes from 8.710 for $N_{Filt}=64$ to 8.634 for $N_{Filt}=16$, while the maximum BER gain remains comparable. This indicates that a relatively small number of filters is sufficient to capture the decision regions for higher-order modulation. However, reducing the model to $N_{Filt}=4$ leads to a significant degradation, with the average BER gain dropping to 1.692 and the minimum gain reducing substantially.


From a complexity perspective, reducing the model size from $N_{Filt}=64$ to $N_{Filt}=16$ yields over $93\%$ reduction in both parameters and FLOPs for both modulation schemes. Further reduction to $N_{Filt}=8$ provides savings above $98\%$, but with noticeable performance degradation, particularly for 16QAM.


These results indicate that $N_{Filt}=16$ offers a strong balance between complexity and performance, achieving substantial reductions in model complexity with minimal loss in accuracy, while $N_{Filt}=8$ remains a viable option for highly resource-constrained deployments. Fig.~\ref{fig:ablation} shows the BER versus SINR performance for models with $N_{Filt}=\{64,16,8,4\}$ for both 16QAM and 64QAM.


\subsection{Generalization to Unseen Environments}

\begin{figure}[b]
    \centering
    \includegraphics[width=\linewidth]{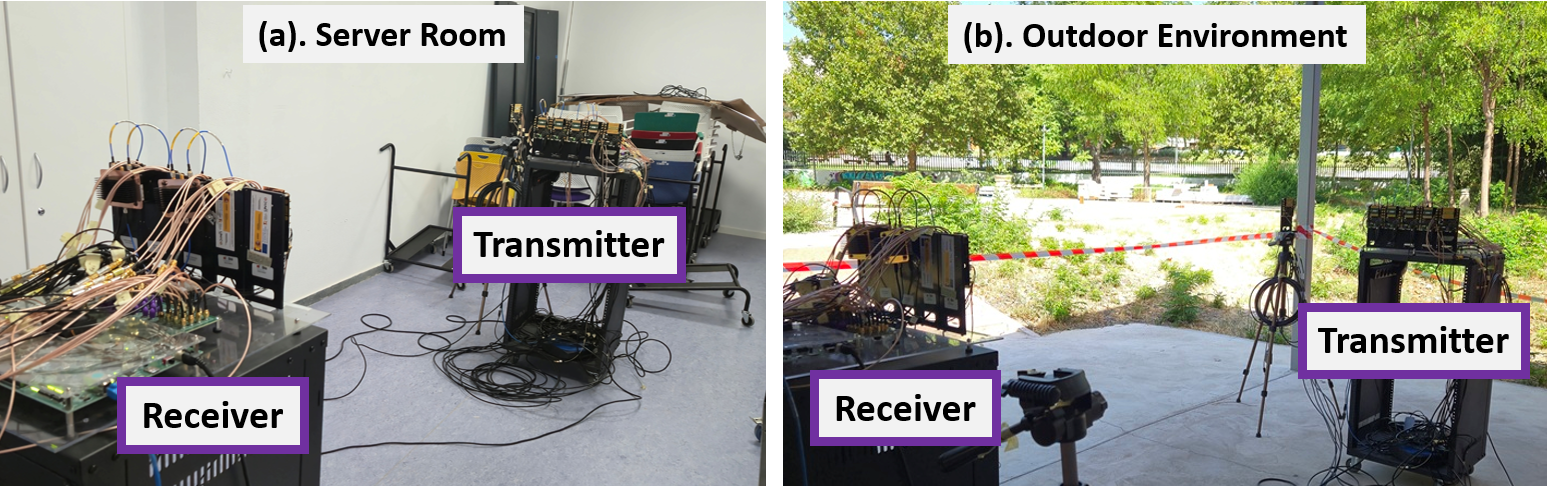}
    \caption{\small Unseen evaluation environments used to assess generalisation of the proposed demapper: (a) outdoor open-space deployment and (b) indoor server-room environment.}
    \label{unseen}
\end{figure}

\begin{figure}[]
    \centering
    \includegraphics[width=0.9\linewidth]{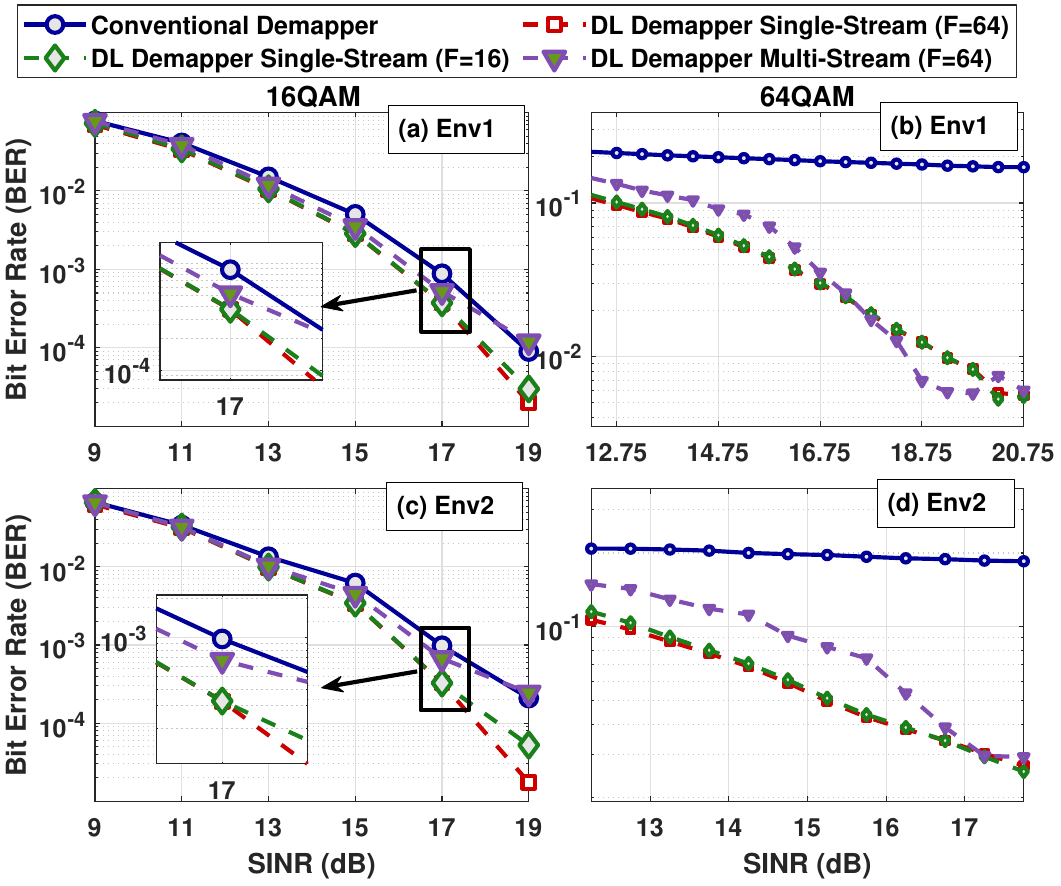}
    \vspace{-2mm}\caption{\small Performance of trained models in unseen environments. 
    Env1: Indoor environment; Env2: Outdoor environment.}
    \label{fig:generalization}
    \vspace{-2mm}
\end{figure}

A fundamental requirement for any learning-based physical layer component is the ability to operate reliably under channel conditions that differ from those observed during training. While the earlier evaluations use training and test datasets collected on different days and at different times, all measurements originate from the same laboratory environment. Although this introduces natural variability in hardware state and channel realization, it does not expose the model to fundamentally different propagation scenarios.

To evaluate generalization, the trained demappers are tested in two unseen environments with markedly different propagation characteristics, as illustrated in Fig.~\ref{unseen}. The first environment is an indoor server room with a compact layout, where objects such as racks, metallic enclosures, and electronic equipment introduce additional reflections and scattering paths. The second environment is an outdoor open space with a clearer propagation path and significantly fewer nearby reflectors. Neither environment is represented in the training data.

Figure~\ref{fig:generalization} shows the BER performance for both 16QAM and 64QAM in the two unseen environments. For both modulation formats, the single-stream demapper exhibits performance that closely matches its behaviour in the original laboratory setting. Despite the substantial change in propagation conditions, including different reflection characteristics, the BER curves remain largely unchanged. This demonstrates that the proposed demapper generalizes well to environments that were not observed during training and does not rely on environment-specific characteristics of the training data.
For completeness, the multi-stream demapper follows the same qualitative behaviour observed earlier. Its performance remains consistently below that of the single-stream configuration and does not show improved robustness in the unseen environments.

These results suggest that the proposed demapper does not overfit to environment-specific features of the training data. Instead, it learns characteristics of the equalized wideband symbols that remain stable across a range of indoor and outdoor mmWave channels. Such environment invariance is critical for practical deployment, where propagation conditions, scattering profiles, and hardware states can vary significantly over time and location.

\section{Hardware Evaluation}\label{sec:hwResults}

The hardware implementation of the proposed deep learning–based demapper is evaluated on the MIMORPH platform~\cite{mimorph}, which is built around the AMD Xilinx ZCU111 RFSoC evaluation board integrating the Zynq UltraScale+ RFSoC device. This heterogeneous SoC combines a quad-core ARM Cortex-A53 application processing unit running at up to 1.333~GHz with programmable logic (PL), offering approximately 425k LUTs, 851k flip-flops, 4272 DSP slices, and 1080 block RAMs (36~Kb each), along with 80 UltraRAM blocks. These resources make the platform well suited for implementing compute-intensive signal processing accelerators, such as neural network–based demappers, while supporting integrated high-speed RF data converters under realistic area and power constraints.

All results reported in this section are obtained from post-implementation designs using the hardware-software co-design framework described in Section~\ref{sec:hwDesign}. Two metrics are used to evaluate the hardware implementation: execution time and resource utilisation. Execution time measures the total latency required to process one data block of $T=448$ symbols, including both computation and data transfers between DDR memory and on-chip buffers. Resource utilisation captures the hardware cost in terms of LUTs, flip-flops, DSP slices, and block RAM usage. Together, these metrics characterise the latency–area trade-off of the accelerator, since reducing execution time typically requires increased parallelism and higher resource consumption.

Based on the ablation study presented in Section~\ref{sec:ablation}, the hardware implementation adopts a configuration with $N_{FILT} = 16$, which provides a favourable balance between BER performance and hardware complexity. This configuration is therefore used for all hardware measurements reported in the following sections.


\begin{table*}[]
\centering
\caption{\small Post-implementation hardware performance, resource utilisation, and power consumption on AMD Xilinx RFSoC.  }
\renewcommand{\arraystretch}{1.2}
 \resizebox{0.9\textwidth}{!}{
\begin{tabular}{|l|cc|cccc|c|}
\hline
\multirow{2}{*}{\textbf{Configuration}} & \multicolumn{2}{c|}{\textbf{Latency}} & \multicolumn{4}{c|}{\textbf{Resource Utilization}} & \multirow{2}{*}{\textbf{\begin{tabular}[c]{@{}c@{}}PL Power\\ (W)\end{tabular}}} \\ \cline{2-7}
 & \multicolumn{1}{c|}{\textbf{\begin{tabular}[c]{@{}c@{}}Execution Time (ms/block)\end{tabular}}} & \multicolumn{1}{l|}{\textbf{Speedup}} & \multicolumn{1}{c|}{\textbf{LUT}} & \multicolumn{1}{c|}{\textbf{FF}} & \multicolumn{1}{c|}{\textbf{BRAM}} & \textbf{DSP} &  \\ \hline
\textbf{PS only} & \multicolumn{1}{c|}{17.5} & 1 & \multicolumn{1}{c|}{-} & \multicolumn{1}{c|}{-} & \multicolumn{1}{c|}{-} & - & - \\ \hline
\textbf{SPFL} & \multicolumn{1}{c|}{0.95} & 18.42 & \multicolumn{1}{c|}{11567} & \multicolumn{1}{c|}{19044} & \multicolumn{1}{c|}{34.5} & 70 & 1.771 \\ \hline
\textbf{FP\textless{}14,6\textgreater{}} & \multicolumn{1}{c|}{0.92} & 19.02 & \multicolumn{1}{c|}{7843} & \multicolumn{1}{c|}{9781} & \multicolumn{1}{c|}{18} & 20 & 1.442 \\ \hline
\textbf{\begin{tabular}[c]{@{}l@{}}FP\textless{}14,6\textgreater  (2 Rx in parallel)\end{tabular}} & \multicolumn{1}{c|}{0.51} & 34.31 & \multicolumn{1}{c|}{13470} & \multicolumn{1}{c|}{17919} & \multicolumn{1}{c|}{36} & 40 & 1.688 \\ \hline
\end{tabular}
}
\label{tab:hw_results}
\end{table*}

\subsection{Floating-Point Software and Hardware}


We begin with the baseline software implementation executed entirely on the PS, whose execution time is reported in Table~\ref{tab:hw_results}. When executed on the ARM cores alone, the demapper requires 17.5~ms per data block, corresponding to approximately 39~$\mu$s per symbol for a block size of 448 symbols. This latency is dominated by the convolution operations, which are computationally intensive in software.

We next evaluate the implementation on the PL using a single-precision floating-point (SPFL) design, operating at a clock frequency of 300~MHz. Since the arithmetic precision is unchanged, the BER performance is identical to that of the PS implementation. As shown in Table~\ref{tab:hw_results}, the SPFL accelerator utilises 11567 LUTs, 19044 flip-flops, 34.5 BRAMs, and 70 DSP slices, corresponding to 2.7\%, 2.2\%, 3.2\%, and 1.6\% of the available resources on ZCU111, respectively. The execution time is reduced to 0.95 ms per block, which corresponds to approximately 2.1~$\mu$s per symbol and represents an 18.4$\times$ speedup compared to the PS implementation. This result confirms that convolution-dominated workloads benefit substantially from hardware acceleration, even when floating-point arithmetic is retained.

These results show that offloading the demapper to the PL provides a substantial latency reduction without compromising detection performance. However, the floating-point implementation still incurs a heavy resource and power cost, and such precision is often unnecessary in practical receivers. The resolution of the analog-to-digital converters (ADCs) limits the number of effective bits in the received samples. This limits the numerical precision that can be meaningfully exploited during processing. As a result, using full floating-point precision does not necessarily provide additional performance gains. This motivates the use of reduced-precision designs discussed next.

\subsection{Word length optimization}


Reducing numerical precision is a key step in translating deep learning models into efficient hardware implementations. In hardware accelerators, arithmetic precision directly impacts the cost of multipliers, adders, memory and data movement, making fixed-point arithmetic preferable to floating-point arithmetic on resource-constrained platforms. A fixed-point representation is denoted by $\langle W, I \rangle$, where $W$ is the total word length and $I$ is the number of integer bits, including the sign bit. Reducing $W$ decreases arithmetic and memory cost, while reducing $I$ limits dynamic range and increases the risk of overflow. Although neural networks are generally robust to moderate quantisation, aggressive precision reduction can introduce accumulated numerical errors and degrade demapping performance. Word-length optimisation therefore requires balancing hardware efficiency and detection accuracy.

In the proposed hardware architecture, a single convolution engine is reused serially across all network layers. As a result, a uniform fixed-point format must be applied consistently to feature maps, weights, and intermediate accumulations. Supporting layer-specific precisions would require additional datapath logic and control complexity, which would undermine the efficiency benefits of the serial design. The use of a single $\langle W, I \rangle$ format is therefore an intentional architectural constraint that simplifies the hardware but places greater emphasis on the careful selection of precision. The fixed-point format is chosen through a two-stage procedure, with performance evaluated against the SPFL baseline using BER as the primary metric.

\begin{figure}[!b]
    \centering
    \includegraphics[width=0.85\linewidth]{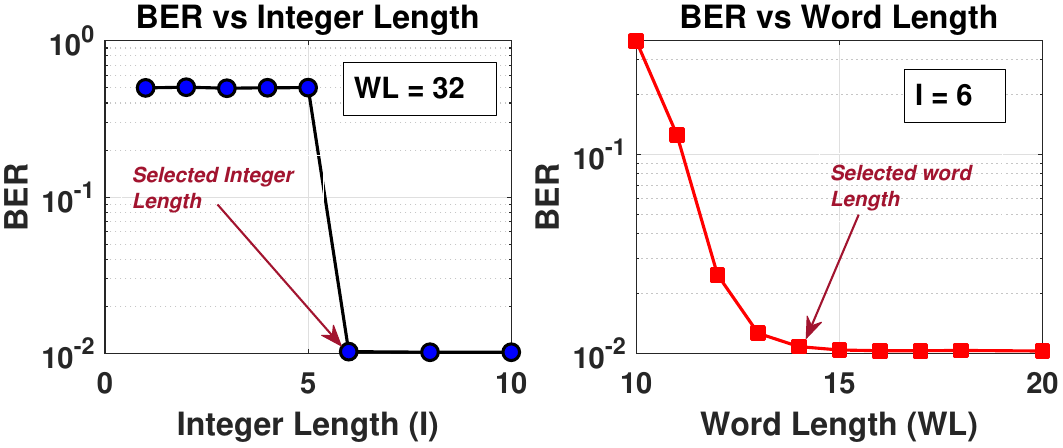}
    \vspace{-2mm}
    \caption{\small Word length analysis for the DL based demapper.}
    \label{fig:wlopt}
\end{figure}

In the first stage, the integer length $I$ is determined by fixing a sufficiently large total word length $W$ and starting with $I = 1$, corresponding to a sign bit with no additional integer bits. The demapper performance is evaluated, and $I$ is incrementally increased until the BER stabilises and no longer deviates significantly from the floating-point baseline, indicating sufficient dynamic range. Once the integer length is fixed, the second stage reduces the total word length $W$ while keeping $I$ constant. Starting from a large value, $W$ is decreased until quantisation noise causes unacceptable BER degradation. The smallest value of $W$ that preserves acceptable performance is then selected.

The results of this analysis are shown in Fig.~\ref{fig:wlopt}. Integer lengths below $I = 6$ result in a sharp increase in BER, indicating insufficient dynamic range, and therefore $I = 6$ is selected. With this integer length fixed, total word lengths below $W = 14$ lead to significant performance degradation. Based on these observations, the final fixed-point format is chosen as $\langle 14, 6 \rangle$, which provides a favourable balance between numerical accuracy and hardware efficiency.

After selecting the fixed-point format $\langle 14, 6 \rangle$, the word-length-optimised demapper is implemented on the PL and evaluated against the SPFL baseline, with the resulting hardware metrics summarised in Table~\ref{tab:hw_results}. The fixed-point implementation utilises 7843 LUTs, 9781 flip-flops, 18 BRAM blocks, and 20 DSP slices, corresponding to 1.8\%, 1.2\%, 1.7\%, and 0.5\% of the available resources on the ZCU111, respectively. Compared to the SPFL design, this represents reductions of 32.2\% in LUT usage, 48.6\% in flip-flop usage, 48.8\% in BRAM usage, and 71.4\% in DSP usage, along with a 18.6\% reduction in PL power. In addition to these resource savings, the execution time per block decreases from 0.95 ms to 0.92 ms due to lower arithmetic complexity.
Furthermore, by instantiating two fixed-point demapper accelerators in parallel, the overall resource utilisation becomes comparable to the single-accelerator floating-point design, while reducing the execution time to 0.51 ms per block, corresponding to a 34.31$\times$ speedup over the PS-only implementation. These efficiency gains are accompanied by a small increase in BER, from 0.0101 to 0.0108 for a representative dataset, which remains limited and acceptable given the substantial reductions in area and power. Overall, the fixed-point implementation achieves a strong balance between demapping performance and hardware efficiency, making it well-suited for resource-constrained mmWave receiver platforms.

\section{Conclusion}\label{sec:conclusion}

This work presented a deep learning–based demapper that performs block-wise processing of equalised symbol sequences while preserving the modular structure of conventional receivers. The proposed approach exploits localized dependencies that are not captured by classical symbol-wise demappers and applies to both single-carrier and multi-carrier systems. The demapper was validated using a newly generated over-the-air dataset collected from a wideband mmWave MIMO system operating at 60~GHz across multiple days and environments. Experimental results showed consistent improvements over a conventional soft-decision demapper, achieving up to a $2.6\times$ reduction in BER for 16QAM and up to a $17.6\times$ reduction for 64QAM. Ablation studies confirmed that the gains primarily arise from block-wise processing. The proposed architecture also generalized well to unseen environments without retraining and achieved up to a $34\times$ speedup on a 
RFSoC implementation using fixed-point arithmetic.

Future work will investigate scalability to higher-order MIMO configurations and evaluate performance under mobility and rapidly time-varying channel conditions. These directions are important for assessing the applicability of learning-based demapping in future 6G wireless systems.

\section*{Acknowledgments}

This work is supported by the Horizon Europe Research and Innovation programme (SNS JU) under Grant Agreements No.~101292662 (IoT-ZERO) and No.~101192521 (MultiX), and by the Comunidad de Madrid through the DISCO6G-CM (TEC-2024/COM-360) and TUCAN6-CM (TEC-2024/COM-460) projects.


\bibliographystyle{IEEEtran}
\bibliography{IEEEabrv,ref}

\end{document}